\documentclass[a4paper,11pt]{article}
\usepackage{pos}
\usepackage{wrapfig}
\usepackage{adjustbox}
\usepackage{physics}
\usepackage{lineno}

\begin{document}

\title{IceCube’s Sensitivity Prospects to MeV-Scale Axion-Like Particles from Core-Collapse Supernovae}
\ShortTitle{IceCube’s Sensitivity Axion-Like Particles from Supernovae}


\author*[a]{Nora Valtonen-Mattila}
\author[b]{Shlok Shah}
\author[b]{Segev BenZvi}

\affiliation[a]{Fakultät für Physik \& Astronomie, Ruhr-Universität Bochum, D-44780 Bochum, Germany}
\affiliation[b]{Department of Physics and Astronomy, University of Rochester\\Rochester, NY 14627, USA}

\emailAdd{nvalto@astro.ruhr-uni-bochum.de}

\abstract{We present a novel framework to estimate the sensitivity and discovery potential of IceCube to axion-like particles (ALPs) produced in core-collapse supernovae (CCSNe), covering ALP masses from 1 MeV to several hundred MeV. A key feature of this work is the explicit handling of the final-state leptons produced in ALP interactions with $^{16}$O nuclei and protons, which can generate Cherenkov light detectable in IceCube. These processes are being fully integrated into a detector-level simulation chain, enabling realistic detector signal modeling beyond existing estimates. The framework enables sensitivity forecasts for both direct detection and constraints based on time delays relative to the neutrino burst, across a range of ALP emission models. This approach may also extend to other MeV-scale dark sector particles. Preliminary sensitivity estimates are in progress and will be presented.

\ConferenceLogo{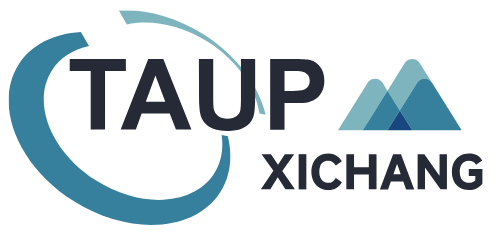}

\FullConference{%
XIX International Conference on Topics in Astroparticle and Underground Physics 2025 (TAUP 2025)\\
  24 August -- 30 August, 2025\\
  Xichang, China}
}

\maketitle

\section{Introduction}\label{sec:Introduction}
Axion-like particles (ALPs) are light pseudoscalar bosons that arise in many extensions of the Standard Model~\cite{DoddyALPs}, including theories motivated by the strong CP problem~\cite{StrongCP} and string compactification \cite{Jaeckel:2010ni}. Their masses and couplings are largely independent, allowing for possible interactions with nucleons, photons, and electrons. In this work, we focus on ALPs that couple to nucleons.

\medskip 

\begin{wrapfigure}[21]{r}{0.55\textwidth}
    \includegraphics[width=\linewidth]{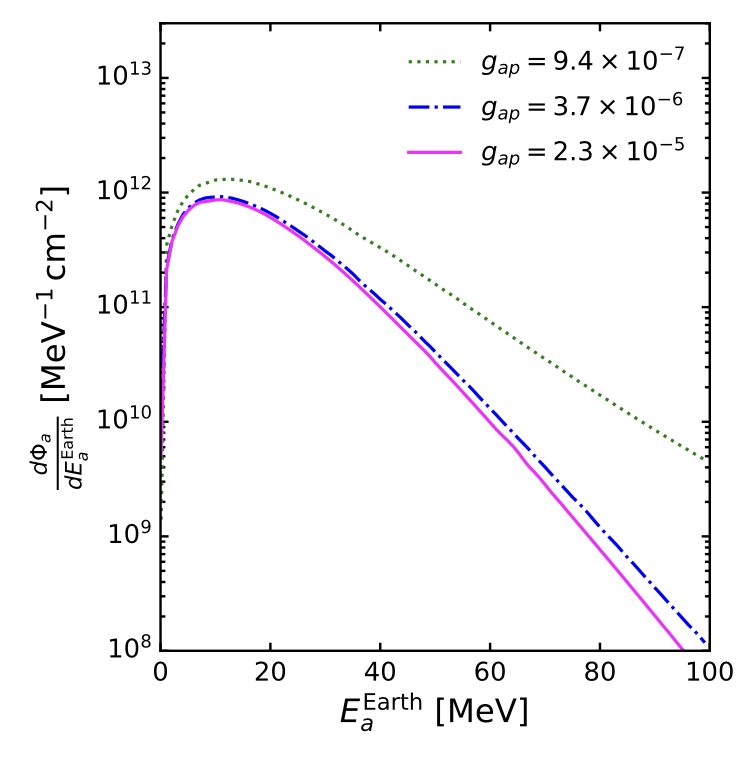}
    \caption[font=large]{Axion-like particle flux from a Core-collapse Supernova at a distance of 1 kpc for different coupling strengths. Reproduced from \cite{AlonsoDelayedSignal}.}
    \vspace{-7pt}
    \label{fig:alp_flux}
\end{wrapfigure}

Core-collapse supernovae (CCSNe) are efficient environments for producing such ALPs, with fluxes comparable to thermal neutrinos~\cite{LellaSNaxions, LellaMassiveALPsBounds} (see Fig.~\ref{fig:alp_flux}). During the collapse and early cooling phase, nuclear densities and temperatures of tens of MeV enable production primarily through nucleon–nucleon bremsstrahlung and related hadronic processes~\cite{CarenzaCheatSheet}. If the coupling is sufficiently weak, the ALPs escape freely and carry away energy, which has been used to constrain ALP–nucleon couplings from the duration of the neutrino burst observed from SN1987A~\cite{Raffelt1987A}.

\medskip

ALPs with couplings to photons or electrons could also lead to complementary signatures, including possible gamma-ray conversion in Galactic magnetic fields or modifications of the neutrino cooling channel, leading to a decrease in neutrino signal~\cite{Takata2025, MoriMultimessenger}. These are indirect search channels. In contrast, the present work focuses on the direct detection of ALPs via their interactions in the detector itself, specifically through scattering on oxygen nuclei~\cite{Carenzaxs} and free protons~\cite{AlonsoDelayedSignal, AlonsoDiffuseFlux}.

\medskip

Large water Cherenkov neutrino observatories such as IceCube are well-suited for this purpose. IceCube detects supernova neutrinos through a collective rise in photomultiplier rates due to positrons from IBD~\cite{IceCube:2011cwc, IceCube:2016zyt}, but MeV-scale ALP interactions can also produce final-state electrons and positrons that generate Cherenkov light in the detector. Furthermore, because ALPs may be mildly relativistic, depending on their mass, their arrival time can differ from that of the prompt neutrino burst. This enables two complementary observables: a direct Cherenkov light signal from ALP interactions, and timing-based constraints based on relative delays.

\medskip 

IceCube has a live time >99\% and is sensitive to supernova neutrinos from the Milky Way \cite{IceCube:2011cwc, IceCube:2016zyt, IceCube:2023ogt}. Given the likelihood of IceCube operating during the next Galactic CCSN, it is timely to develop realistic, detector-level sensitivity projections for nucleon-coupled ALPs. In this work, we present a simulation framework that integrates ALP production models, propagation, interaction cross sections, and full detector response, enabling studies of both direct detection potential and timing signatures across a broad range of ALP masses.

\section{ALPs from CCSNe}\label{sec:ALPs}

Core-collapse supernovae provide a dense and hot environment where ALPs can be efficiently produced through strongly interacting processes. In the proto-neutron star (PNS), typical temperatures reach T$\sim$30–50 MeV \cite{Fischer:2021tvv} and densities approach or exceed nuclear saturation. Under these conditions, ALP production is dominated by interactions involving nucleons and pions. The two main channels by which these ALPs can be produced are nucleon–nucleon bremsstrahlung, $N+N \rightarrow N + N + a$~\cite{ALPBrehm}, and pion-induced processes like $\pi^- + p \rightarrow n + a$~\cite{ALPCompPion}.

\medskip
The escape or trapping of ALPs depends on the strength of the ALP-nucleon coupling, $g_{aN}$. For sufficiently small $g_{aN}$, ALPs free-stream out of the proto–neutron star (PNS) and efficiently carry away energy. As $g_{aN}$ increases, ALPs begin to scatter more frequently on nucleons and the mean free path decreases, reducing the emitted luminosity. For even larger couplings, ALPs become trapped and eventually thermalize with the medium. This transition is continuous, leading to a characteristic dependence of the ALP emission on both the mass and the coupling. Constraints from the duration of the SN1987A neutrino burst disfavor the region in which ALPs are produced abundantly and escape freely, since this would lead to excessive cooling of the proto–neutron star \cite{Raffelt:1996wa}. At larger couplings, where ALPs become trapped and thermalize, this constraint weakens again \cite{Raffelt:1987yt, Raffelt:1996wa}. As a result, a significant range of ALP masses and nucleon couplings remains unconstrained.

\medskip

For masses in the MeV-to-hundreds-of-MeV range, the kinematics of the production processes modify the outgoing spectrum and the velocity distribution of ALPs arriving at Earth. In particular, at higher masses, the population of ALPs becomes less relativistic, which can lead to measurable arrival time delays relative to the prompt neutrino burst if the ALPs interact in a terrestrial detector. These delays provide an additional observable that can be exploited in combination with direct detection signatures, as discussed in Section~\ref{sec:detection_system}.

\section{ALP Direct Detection}\label{sec:detection_system}

This work focuses on the direct detection of nucleon–coupled ALPs in IceCube. For nucleon couplings, indirect constraints arise from the effect of ALP emission on the cooling of the proto–neutron star, which would modify the duration of the neutrino burst observed from SN1987A. Meanwhile, for photon-coupled ALPs, indirect searches are based on ALP–photon conversion into gamma-rays in Galactic magnetic fields \cite{Manzari:2024jns}. Here, we focus solely on the nucleon-coupling scenario and its direct detection signature in IceCube.

Given an ALP flux at Earth, we include the interactions
\begin{align}
a + p &\rightarrow p + \gamma, \
&
a + {}^{16}\mathrm{O} \rightarrow {}^{16*}\mathrm{O}^{} \rightarrow {}^{16}\mathrm{O} + \gamma .
\end{align}
The ALP–proton interaction proceeds as a radiative scattering process, producing a $\gamma$-ray in the final state. In contrast, interactions with ${}^{16}\mathrm{O}$ can excite nuclear levels, followed by de-excitation with the emission of a $\gamma$-ray at characteristic discrete energies. The relative contributions of proton and oxygen targets are determined by the corresponding excitation and scattering cross sections and can be varied to reflect model uncertainties.

\medskip

For ${}^{16}\mathrm{O}$ targets, the interaction can populate excited nuclear states, which subsequently decay through cascades of $\gamma$-ray emission determined by the nuclear energy levels and the branching ratios. For proton targets, the ALP interaction proceeds as a radiative scattering process, producing a continuous-$\gamma$-ray spectrum set by the differential scattering cross section. We simulate the production of gamma rays from both interactions, and in both cases, the resulting $\gamma$-rays are propagated through the ice using a dedicated Monte Carlo simulation that treats photoelectric absorption, Compton scattering, and pair production, with bremsstrahlung included for high-energy electrons. The cascade is followed until particles fall below the critical energy in ice, yielding the final $e^\pm$ spectra used as inputs to the detector response model \cite{ASTERIA-zenodo:2025, ASTERIA-readthedocs:2025}.

\subsection{Detector Response}

The lepton spectra are folded through a fast detector response model specifically developed for this work. This model converts lepton energy and geometry into expected digital optical module (DOM) hit counts by incorporating the Cherenkov photon yield, DOM angular acceptance, noise and triggering characteristics of IceCube's Supernova data stream \cite{IceCube:2011cwc}. Since IceCube’s supernova detection channel operates as a rate-based measurement rather than event-by-event reconstruction~\cite{IceCube:2011cwc, ASTERIA-zenodo:2025}, the final observable is the excess in total DOM hit rates over the burst timescale, typically $\sim$10 seconds. The method therefore provides sensitivity to ALP-induced collective rate increases coincident with a core-collapse neutrino burst. This framework provides a direct-search channel for nucleon-coupled ALPs in IceCube, mapping ALP fluxes to expected detector-rate signatures through nuclear excitation, $\gamma$-ray production, secondary lepton cascades, and detector-response modeling.

\section{Detection Horizon} 

\begin{figure}[h]
\centering
    \vspace{-\baselineskip} 
    \includegraphics[width=0.8\linewidth]{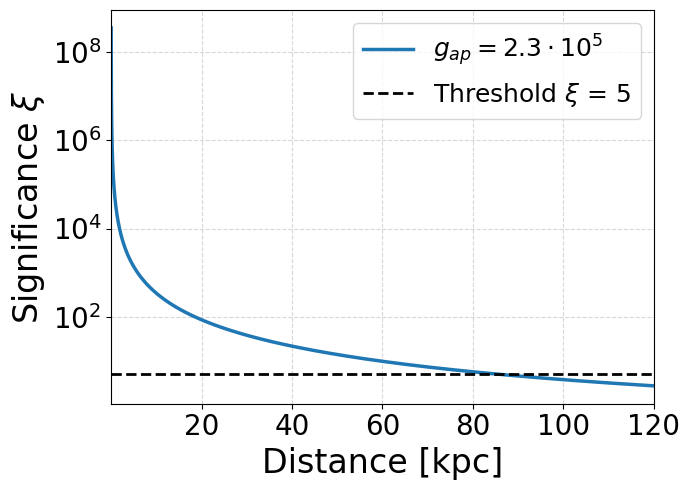}
    \caption[font=large]{IceCube detection significance as a function of distance for an ALP flux like that of Fig.~\ref{fig:alp_flux} for the O16 interaction channel. We set a cut-off of $\xi=5$, which can be approximated to 5$\sigma$.}
    \label{fig:detection_significance}
\end{figure}

With the simulation framework described above, we can evaluate the expected detector response to ALPs from a core-collapse supernova as a function of the ALP mass, coupling strength, and supernova distance. While the full sensitivity study is still in progress, preliminary results indicate that, for benchmark nucleon couplings consistent with existing constraints, the resulting signal would be detectable for a supernova occurring anywhere within the Milky Way, LMC and SMC, as can be seen in Fig. \ref{fig:detection_significance} for one ALP model with $g_{ap}=2.3 \cdot 10^{-5}$ and mass $m_a \ll 1\,  \text{MeV}$ \cite{AlonsoDelayedSignal}. A comprehensive evaluation of the detection horizon and exclusion potential for the direct detection of ALPs with IceCube will be presented in a future publication.

\section{Conclusions}\label{sec:conclusions}

We have introduced a framework to evaluate the sensitivity of IceCube to axion-like particles produced in core-collapse supernovae, focusing on ALPs that couple to nucleons. The approach incorporates ALP production models, nuclear excitation interactions in ice, secondary $\gamma$-ray and lepton propagation, and detector-level response, enabling a direct search strategy complementary to existing indirect constraints. Preliminary studies indicate that IceCube has the potential to probe ALP signals from supernovae across the Milky Way. Final sensitivity results and detection reach will be presented in a forthcoming work.

\bibliographystyle{ICRC}
\bibliography{references}

\clearpage
%
%
%


\end{document}